\documentclass{jfm}

\usepackage{graphicx}
\usepackage{natbib}
\usepackage{amsmath}
\usepackage{hyperref}
\usepackage{stackengine}
\usepackage{color}
\usepackage{comment}

\title[Spatio-temporal spectra in wall turbulence]{Spatio-temporal spectra in the logarithmic layer of wall turbulence: large-eddy simulations and simple models}

\author[Michael Wilczek, Richard J.A.M. Stevens and Charles Meneveau]
{Michael Wilczek$^{1,2}$, Richard J.A.M. Stevens$^{1,3}$ and Charles Meneveau$^1$}

\affiliation{
$^1$Department of Mechanical Engineering, Johns Hopkins University, Baltimore, Maryland 21218, USA.\\
$^2$Max Planck Institute for Dynamics and Self-Organization, D-37077 G\"ottingen, Germany.\\
$^3$Department of Science and Technology and J.M. Burgers Center for Fluid Dynamics, University of Twente, P.O. Box 217, 7500 AE Enschede, The Netherlands.}

\newcommand{\bs}{\boldsymbol}

\begin{document}

\maketitle

\begin{abstract}

Motivated by the need to characterize the spatio-temporal structure of turbulence in wall-bounded flows, we study wavenumber-frequency spectra of the streamwise velocity component based on large-eddy simulation (LES) data. The LES data are used to measure spectra as a function of the two wall-parallel wavenumbers and the frequency in the equilibrium (logarithmic) layer.  We then reformulate one of the simplest models that is able to reproduce the observations: the random sweeping model with a Gaussian large-scale fluctuating velocity and with additional mean flow. Comparison with LES data shows that the model captures the observed temporal decorrelation, which is related to the Doppler broadening of frequencies. We furthermore introduce a parameterization for the entire wavenumber-frequency spectrum $E_{11}(k_1,k_2,\omega;z)$, where $k_1$, $k_2$ are the streamwise and spanwise wavenumbers, $\omega$ is the frequency and $z$ is the distance to the wall. The results are found to be in good agreement with LES data.
\end{abstract}

\section{Introduction}\label{introduction}

Characterization of the spatio-temporal structure of wall-bounded turbulence at high Reynolds numbers is important from both a fundamental and an applied point of view. 
From a fundamental perspective, one is, for example, interested in the evolution of spatially extended boundary layer flow structures as well as their time evolution (see, e.g., \cite{smits11arf,jimenez13pof} and references therein). The region in which both the mean velocity and the variance exhibit a logarithmic dependence with distance to the wall as (possibly) universal statistical features is of particular interest. From a practical viewpoint, a space-time description of fluctuations is, for example, important to understand correlations of wind-turbine power output at different points and times inside wind farms interacting with the turbulent atmospheric boundary layer. 

Here, we study the space-time correlations of a high-Reynolds-number wall-bounded flow in the logarithmic layer. Spatio-temporal correlations can be equivalently described in Fourier space, which leads us to the study of the wavenumber-frequency ($\bs k$-$\omega$) spectrum in the present work. To capture spatial correlations we are interested in the joint spectrum as a function of both the streamwise and spanwise wavenumbers, in addition to the frequency.

To resolve the joint space-time structure in the logarithmic layer, extensive data are needed to cover the relevant length- and time-scales of the problem as well as to allow for sufficient statistical convergence. We are interested in the logarithmic layer outside the near-wall regions, where viscosity is expected to play an important role. While direct numerical simulations (DNS) would be prohibitive, large-eddy simulations (LES) enable us to resolve a comparably extended logarithmic region at affordable computational cost. In turn, this allows us to accumulate long-time data to capture the temporal structure with sufficient resolution. As the smallest turbulent scales are not resolved by LES, small-scale effects are explicitly neglected (see \cite{he04pof} for analyses of related issues). Because decorrelation in turbulent flows is dominated by large-scale sweeping effects, we anticipate that the inherent limitations of LES will not greatly affect the results, although we must keep these limitations in mind, from the outset.  

In addition to evaluating the spectra from numerical data, we also introduce a model for the $\bs k$-$\omega$ spectrum. While theoretical models for the streamwise wavenumber spectrum exist, for example in terms of the ``attached eddy hypothesis" \citep{perry82jfm} and classical Kolmogorov phenomenology, less emphasis has been placed on including temporal correlations in analytically tractable models. Two major processes affect time correlations of turbulent fluctuations: advection with a mean velocity as well as turbulent advection with large-scale eddies, also known as random sweeping. The latter effect has been proposed as a mechanism for temporal decorrelation by \cite{kraichnan64pof} and \cite{tennekes75jfm} and has been studied in many works ever since. To name only a few, \cite{lumley65pof,wyngaard77jas} and \cite{george89afm} have studied the influence of large-scale flow variation on measured one-dimensional spectra in the context of establishing corrections to Taylor's frozen flow hypothesis. The validity and limitations of the random sweeping hypothesis have been studied,  for example, by \cite{praskovsky93jfm} and by \cite{katul95fdr}. With respect to atmospheric flows, a recent monograph by \cite{wyngaard10book} provides an excellent account of the matter. \cite{chen89pfa} (see also references therein) discuss various theoretical implications of random sweeping decorrelation. Spatio-temporal correlations in the context of wall-bounded flows have been discussed with respect to the question of local convection velocity, for example, by \cite{wills64jfm,fisher64jfm} and more recently by \cite{alamo09jfm}.
Space-time correlations in turbulent shear flows have recently been discussed by \cite{he06pre} and \cite{zhao09pre}, who proposed an elliptical parameterization of space-time correlations of turbulent velocity fluctuations. We note in passing that experimental measurements of the $\bs k$-$\omega$ spectrum have been discussed by \cite{lehew11exf}. A recent review on the history of space-time correlations in turbulent flows is given by \cite{wallace14tam}.

In the derivation of our model we combine some of these ideas to obtain a parameterization for the full $\bs k$-$\omega$ spectrum of the streamwise velocity, $E_{11}(k_1,k_2,\omega;z)$,  where $k_1$, $k_2$ are the streamwise and spanwise wavenumbers, $\omega$ is the frequency and $z$ is the distance to the wall in the equilibrium layer. Extending a recent work by \cite{wilczek12pre}, the model is obtained from a linear advection equation featuring mean flow and large-scale random sweeping advection. We show that the advection model predicts the $\bs k$-$\omega$ spectrum as a product of the wavenumber spectrum and a frequency distribution. To test this prediction, we evaluate the wavenumber spectrum as well as the mean and sweeping velocities from LES data, construct the $\bs k$-$\omega$ spectrum based on the linear advection equation and compare it with the $\bs k$-$\omega$ spectrum obtained from the LES data.

For practical purposes also an analytical model parameterization for the full spectrum is desirable. In the final part of the paper we combine classical asymptotic behavior of spectra from homogeneous isotropic turbulence with parameterizations of boundary layer flows to provide a parameterization of the $\bs k$-$\omega$ spectrum which is then also compared with the data.

\section{Large-eddy simulation results}\label{sec:lesresults}

The results presented in this section are obtained from an LES of (wall-modeled) fully rough-wall turbulent (half) channel flow. In these simulations the mean flow is driven by a constant pressure gradient in the streamwise direction. The computational domain is $L_x/H \times L_y/H \times L_z/H = 4\pi \times 2\pi \times 1$, discretized on a grid with $1024 \times 512 \times 256$ grid points.  As usual, the imposed pressure gradient and the domain height $H$ can be used to define the imposed friction velocity $u_*$, with the pressure gradient equaling $-u_*^2/H$. The flow is fully developed and periodic in both the streamwise and spanwise directions. The LES code uses  pseudo-spectral discretization in the horizontal directions and an energy-conserving second-order finite differencing scheme in the vertical direction \citep{albertson99wrr,porteagel00jfm,bou05}. The imposed roughness scale at the ground is $z_0/H=10^{-4}$, where a standard equilibrium wall model is used to prescribe the wall stress. The subgrid-scale stresses are modeled with the scale-dependent Lagrangian model \citep{bou05}. The presented simulation corresponds to case D2 from \cite{ste14d} where further details about the simulation can be found. The time step has been fixed at $2.5 \times10^{-5} H/u_*$, and data have been gathered during a period of $8.2 \times 10^4$ time steps after the statistically stationary state has been reached. For later post-processing, $8200$ snapshots of the velocity field (i.e. every tenth time step) from five horizontal planes at varying distances from the wall have been stored. To calculate the spectra, windowing has been used in the time-domain. In this paper we present results for a fixed height $z/H \approx 0.154$, but we have checked their validity for a range of different heights within the logarithmic layer, which is the topic of another publication \citep{wilczek15jot}.

\begin{figure}
\begin{center}
  \topinset{(a) \hspace{4.3cm} (b)}{\includegraphics[width=0.515\textwidth]{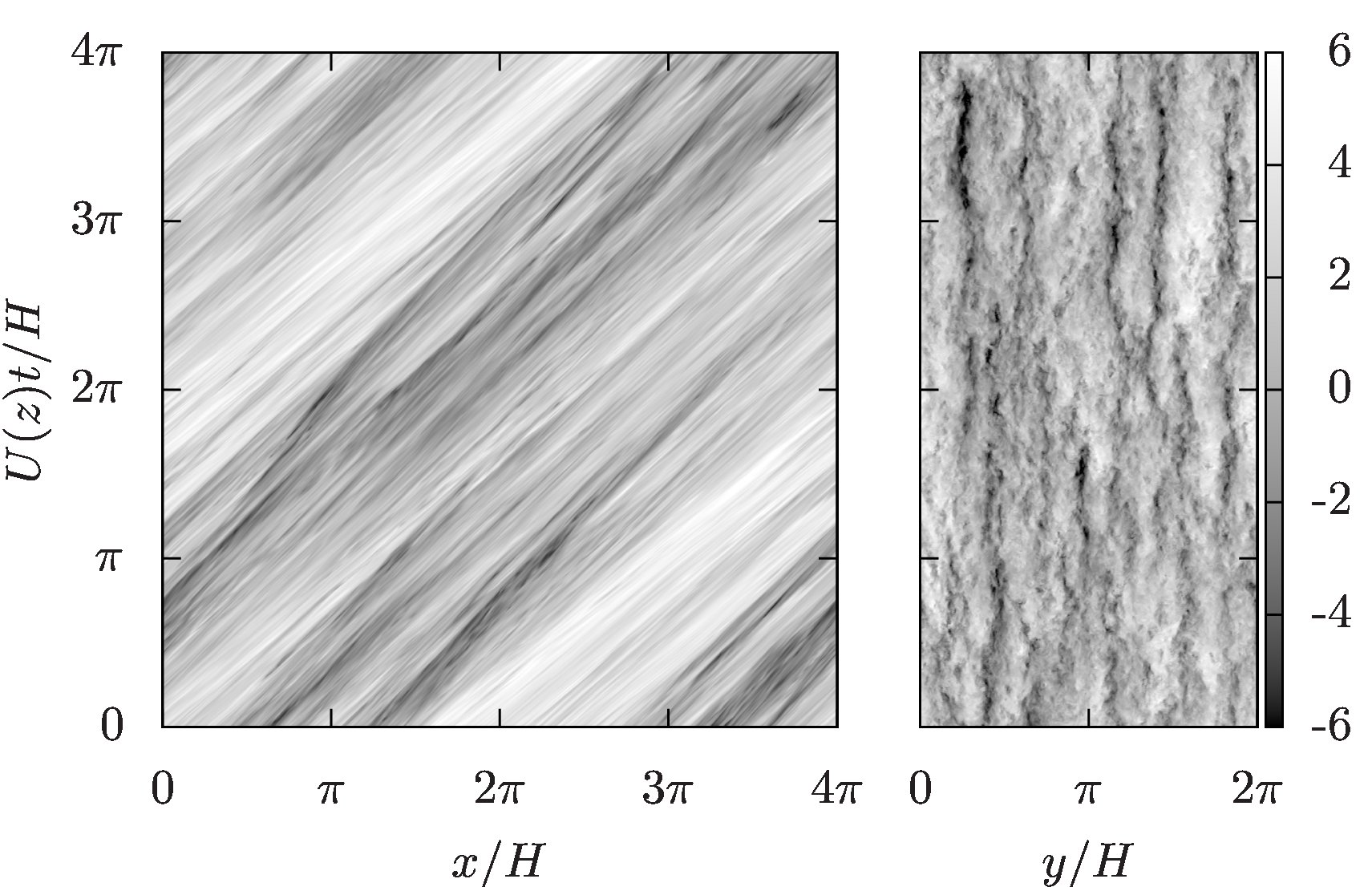}}{0.3cm}{0.1cm}
  \topinset{(c)}{\includegraphics[width=0.47\textwidth]{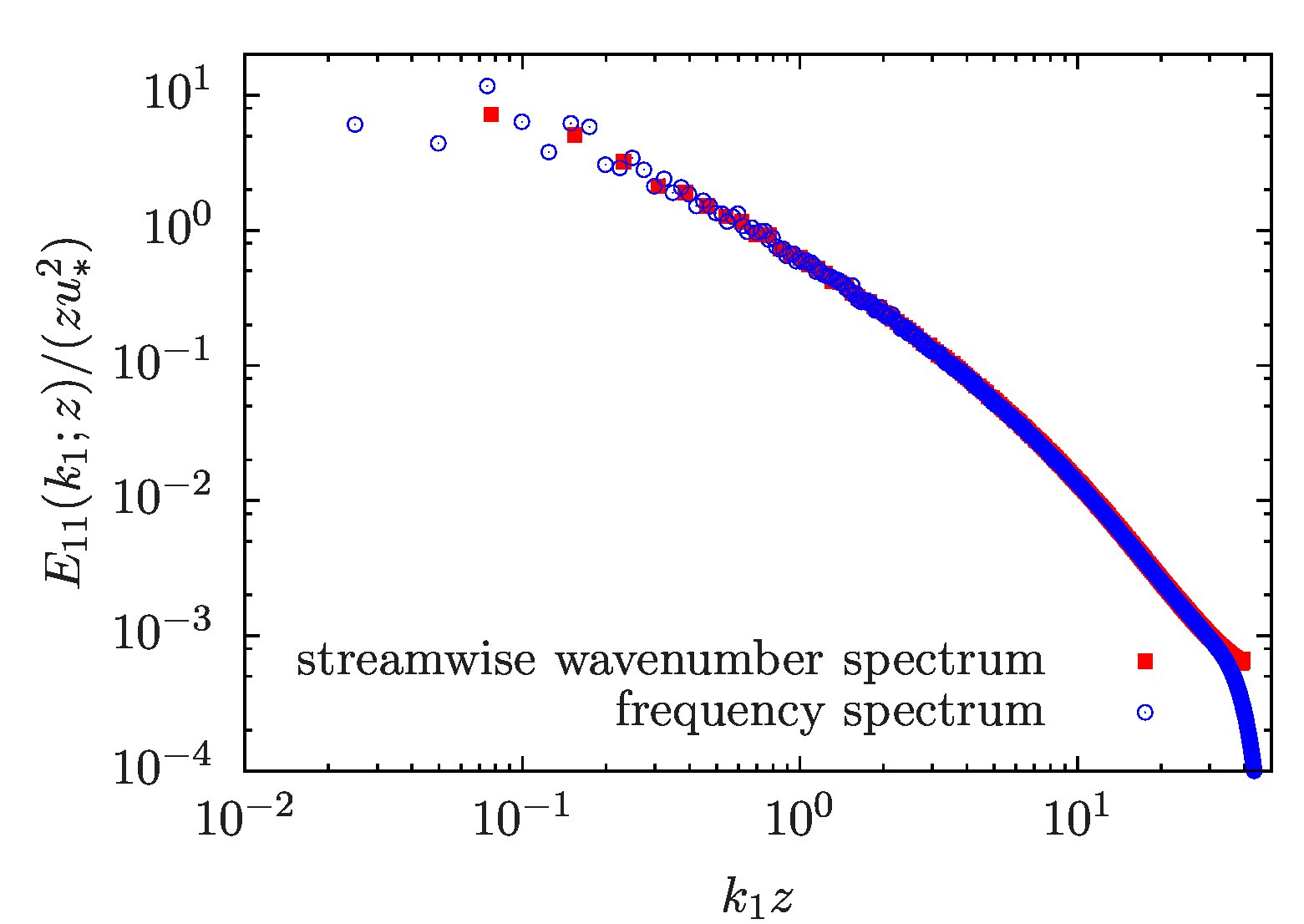}}{0.3cm}{2.5cm}
\end{center}
\caption{Space-time plots of the streamwise velocity fluctuations, along the streamwise (a) and spanwise directions (b) from LES at $z/H\approx 0.154$  (color bar in units of the friction velocity $u_*$). (c) Streamwise wavenumber spectrum of the streamwise velocity component along with the frequency spectrum, interpreted in terms of wavenumber using Taylor's hypothesis ($k_1 = \omega/U(z)$, where $U(z)$ denotes the height-dependent mean velocity).}
\label{fig:spacetime}
\end{figure}

Figure \ref{fig:spacetime} shows space-time plots of the streamwise velocity component, in both the streamwise and spanwise directions. The cut along the streamwise direction (a) clearly shows signatures of mean-flow advection as well as random sweeping effects. Compared with this, the cut along the spanwise direction (b) lacks the advection with the mean flow, which leads to a quite different space-time pattern, dominated by large-scale random sweeping effects as well as the temporal evolution of the small-scale fluctuations. The streamwise wavenumber spectrum of the streamwise velocity component is depicted in figure \ref{fig:spacetime}(c).  Although no clear scaling behavior is observed due to the limited resolution of the LES, the spectrum is not incompatible with two distinct ranges, an approximate $k_1^{-1}$ scaling at low wavenumbers (associated with the log region of the flow below $k_1 z \sim 1$) and an $\sim k_1^{-5/3}$ range at higher wavenumbers. We have also evaluated the frequency spectrum, interpreted as a wavenumber spectrum by means of Taylor's hypothesis for comparison. As expected for a moderate turbulence intensity of approximately $11 \%$ at this height, the frequency spectrum compares well across a broad range of scales. Only close to the high-wavenumber cutoff do differences become apparent, as the pronounced LES wavenumber cutoff is smeared out in the frequency domain. These results indicate that evaluation of the spatial correlation in terms of the $k_1$ spectrum or the temporal correlations in terms of the $\omega$ spectrum (using Taylor's hypothesis) alone yields nearly identical information.

\begin{figure}
\begin{center}
    \topinset{(a) \hspace{2.6cm} (b)}{\includegraphics[width=0.58\textwidth]{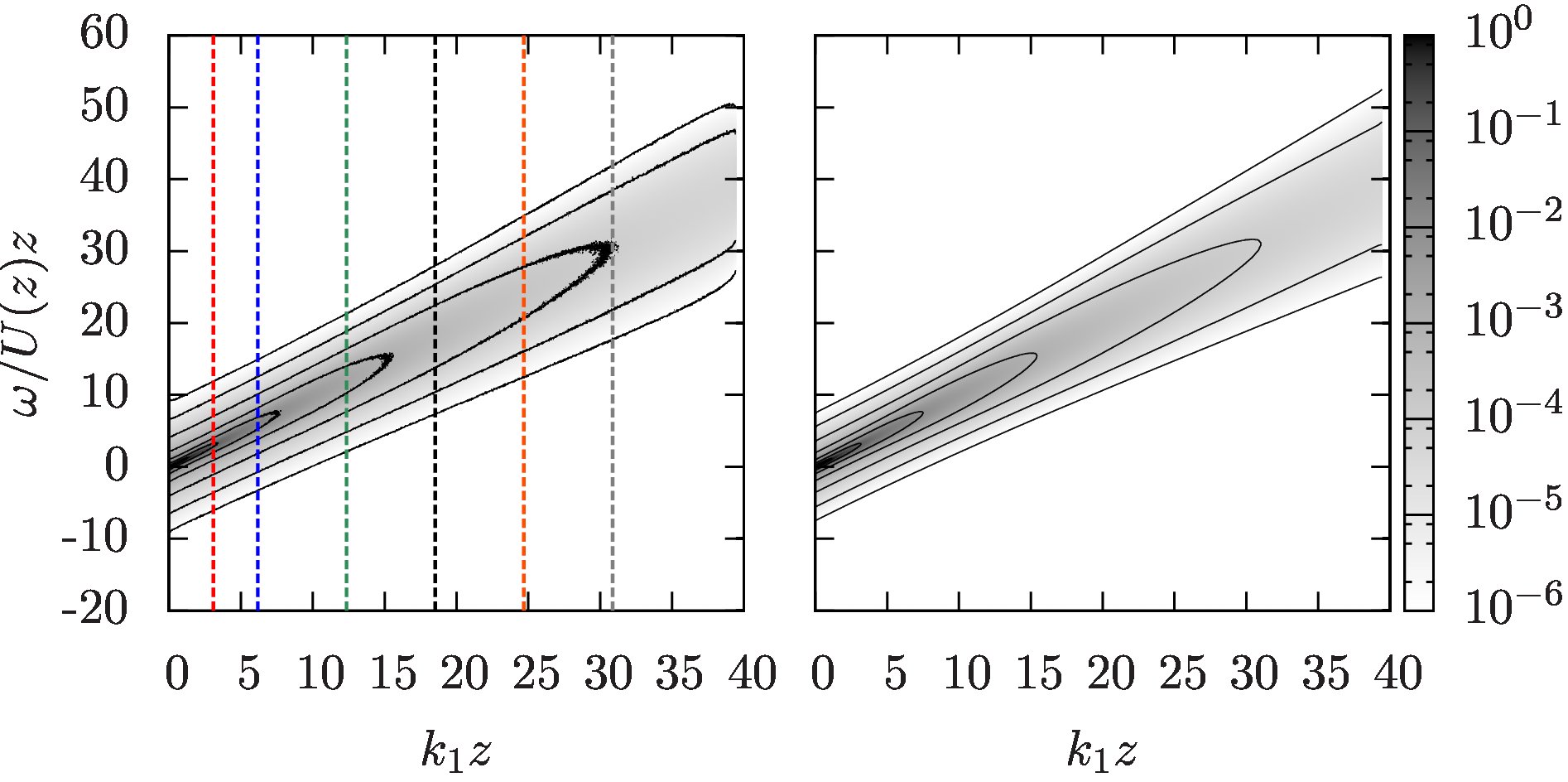}}{2.6cm}{1.1cm}
    \topinset{(c)}{\includegraphics[width=0.41\textwidth]{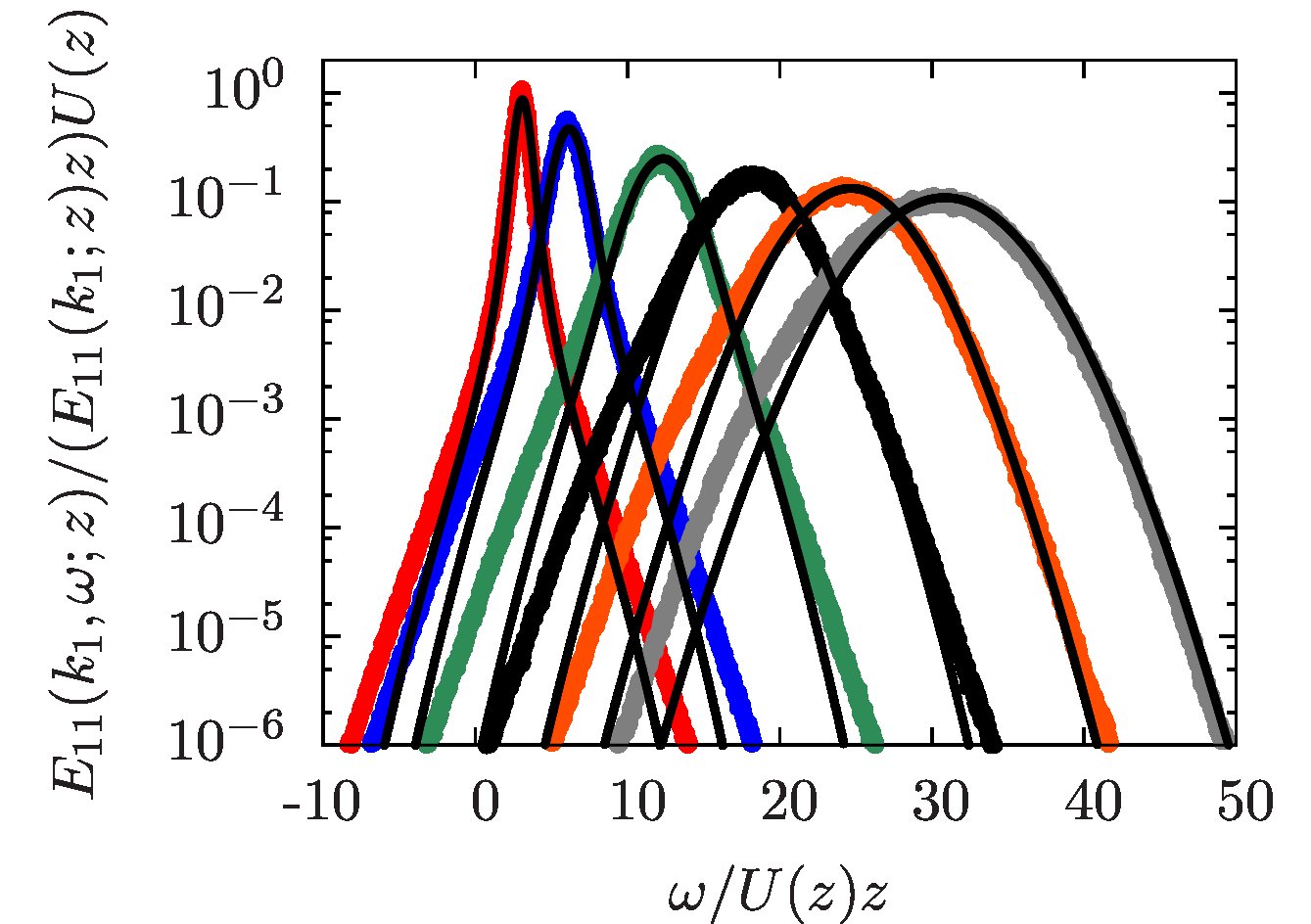}}{0.4cm}{2.1cm}
  \end{center}
\caption{The $k_1$-$\omega$ spectra of the streamwise velocity component at $z/H \approx 0.154$: (a) the spectrum evaluated entirely from LES; (b) the spectrum obtained from \eqref{eq:wavenumberfrequencyspectrum}, for which only the wavenumber spectrum and the mean and random sweeping velocities ($U \approx 19.0 \, u_*$, $\langle v_1^2 \rangle \approx 4.33 \, u_*^2$ and $\langle v_2^2 \rangle \approx 1.70 \, u_*^2$) evaluated from the LES data have been used; (c) normalized cuts through the $k_1$-$\omega$ spectrum from LES (colors), along with the results from the linear random advection model (black lines). The wavenumbers for the cuts are indicated in (a).}
\label{fig:wavenumberfrequencyspectra}
\end{figure}

Further insights can be gained from studying the joint $k_1$-$\omega$ spectra. Figure \ref{fig:wavenumberfrequencyspectra} shows the $k_1$-$\omega$ spectrum of the streamwise velocity component, i.e. the spectrum resolved with respect to the streamwise wavenumber and the frequency. It is obtained by first calculating  the spectrum of a space-time slice of the streamwise velocity component in the streamwise direction. Averaging over the spanwise direction is used to increase the statistical convergence. As can be inferred from figure \ref{fig:wavenumberfrequencyspectra}(a), the spectrum is tilted towards positive frequencies due to the Doppler shift induced by the mean velocity. Moreover, the Doppler broadening due to the random advection is clearly visible. Both effects become more pronounced with increasing wavenumber. It is interesting to note that also the low wavenumbers feature a significant Doppler broadening. This aliasing effect can be attributed to contributions from modes in the spanwise and vertical directions not resolved in this projection. We note that the observed magnitude of Doppler broadening depends on the resolution of the LES; for a better resolved LES higher modes with larger Doppler broadening contribute to the aliasing. From the data sets suited for the current study in terms of stored temporal resolution, we therefore use the one with the highest spatial resolution available.

\section{Linear random advection model}\label{sec:linadvectioneq}

In this section we present a compact rederivation of the linear random advection model. In spite of its very strong inherent assumptions and limitations, it will be shown to provide excellent predictions of the measured $\bs k$-$\omega$ spectrum of the streamwise velocity component $u$ in a horizontal plane. As summarized in section \ref{introduction}, the subject has a rich history, and various formulations for the random sweeping model have been introduced \citep{kraichnan64pof,lumley65pof,tennekes75jfm,wyngaard77jas}.  Additional  background on the presently used formulation is provided by \cite{wilczek12pre}, in which a similar model spectrum has been worked out for homogeneous isotropic turbulence with mean flow.

The starting point of the model derivation is to consider the advection of the streamwise velocity fluctuations $u$, assumed to be statistically homogeneous, in a plane at height $z$ with a mean velocity $\bs U(z) = U(z) \bs e_1$, where $\bs e_1$ is the unit vector pointing in the streamwise direction. Additionally, large-scale random advection by a planar velocity field $\bs v$ with zero mean is included. We make the strong assumption of scale separation between the large-scale velocity field and the fluctuating velocity $u$, such that the random advection velocity can be considered to be approximately constant in space and time compared with the small-scale velocity fluctuations $u$ being advected. Randomness is included by assuming a Gaussian ensemble distribution for $\bs v$ with covariance $\langle v_i v_j \rangle(z)$. With these assumptions we are in first approximation led to a linear advection equation ($u$ is treated as a passive scalar). The equation can be conveniently written in Fourier space for the planar Fourier transform of the velocity fluctuation at a given height $z$: 
\begin{equation}\label{eq:linadvection}
  \frac{\partial}{\partial t} \hat u(\bs k,z,t) + \mathrm{i} (\bs U + \bs v) \cdot \bs k \,  \hat u(\bs k,z,t) = 0 \, ,
\end{equation}
with $\bs k$ representing the streamwise and spanwise wavenumbers, $\bs k = (k_1,k_2)^{\mathrm{T}}$. It should be noted that this equation represents a combination of Taylor's frozen eddy and the Kraichnan-Tennekes random sweeping hypotheses  \citep{taylor38prs,kraichnan64pof,tennekes75jfm}. 

We would also like to stress that the assumption that there are no dynamical interactions between the mean velocity, the random advection and the turbulence cannot be strictly true \citep{praskovsky93jfm,katul95fdr}. It is used here as the simplest possible model for the purpose at hand. Conveniently, advection of $u$ with constant large-scale velocities implies that an initially solenoidal small-scale velocity field remains solenoidal, such that we do not need to explicitly include a pressure term to enforce the divergence-free condition. In this approach we are neglecting the effect of shear on the propagation and sweeping velocities. Such effects could be included, for example, as a shear-enhanced sweeping velocity as proposed by \cite{zhao09pre}.
Solution of \eqref{eq:linadvection} yields
\begin{equation}\label{eq:linadvectionsol}
 \hat u(\bs k,z,t) = \hat u(\bs k,z,0) \exp\left[ -\mathrm{i} (\bs U + \bs v) \cdot \bs k \,  t \right] .
\end{equation}
This result can be used to obtain the two-time covariance of Fourier coefficients. Together with the assumption that $\bs v$ and the inital condition for $u$ are statistically independent we obtain
\begin{equation}\label{eq:fouriercovariance}
  \left\langle \hat u(-\bs k,z,t) \hat u(\bs k,z,t+\tau)  \right\rangle = \left\langle \hat u(-\bs k,z,0) \hat u(\bs k,z,0) \right\rangle \left\langle \exp\left[ -\mathrm{i} (\bs U + \bs v) \cdot \bs k \,  \tau \right] \right\rangle \, .
\end{equation}
This result directly translates into a relation between the instantaneous and the two-time wavenumber spectrum:
\begin{equation}\label{eq:twotimespectrum}
  E_{11}(\bs k,\tau;z) = E_{11}(\bs k;z) \left\langle \exp\left[ -\mathrm{i} (\bs U + \bs v) \cdot \bs k \,  \tau \right] \right\rangle  \, .
\end{equation}
We now evaluate the average on the right-hand side explicitly. The mean velocity $\bs U$ is assumed to be the same across all realizations, whereas we assume a Gaussian ensemble distribution for the large-scale random advection velocity $\bs v$. Under these assumptions, the integration over a Gaussian probability distribution for $\bs v$ can be carried out analytically, and we obtain for the average in \eqref{eq:fouriercovariance} and \eqref{eq:twotimespectrum}
\begin{equation}\label{eq:evaluationofaverage}
  \left\langle \exp\left[ -\mathrm{i} (\bs U + \bs v) \cdot \bs k \,  \tau \right] \right\rangle = \exp\left[ -\mathrm{i} \bs U \cdot \bs k \,  \tau \right] \exp\left[ -\frac{1}{2} \left\langle (\bs v \cdot \bs k)^2  \right\rangle  \tau^2 \right].
\end{equation}
The same term for the random sweeping contribution has, for example, been obtained by \cite{wyngaard77jas}. A Taylor expansion retaining only the first two terms has been studied by \cite{lumley65pof}. 
Here we take the final step to obtain the $\bs k$-$\omega$ spectrum by an additional Fourier transform into frequency space, yielding
\begin{equation}\label{eq:wavenumberfrequencyspectrum}
  E_{11}(\bs k,\omega;z) = E_{11}(\bs k;z) \left[2\pi \left\langle (\bs v \cdot \bs k)^2  \right\rangle \right]^{-1/2} \exp\left[ -\frac{(\omega - \bs k \cdot \bs U)^2}{2 \left\langle (\bs v \cdot \bs k)^2  \right\rangle}  \right] \, .
\end{equation}
In the framework of the linear advection equation we thus obtain the $\bs k$-$\omega$ spectrum as a product of the wavenumber spectrum with a Gaussian frequency distribution. The mean of this distribution is parameterized by the mean velocity, whereas the variance is related to the covariance of the large-scale random advection velocity. The mean velocity leads to a Doppler shift of frequencies, whereas the random advection results in a Doppler broadening. We note in passing that the Gaussian frequency contribution is a direct consequence of the quadratic $\tau$ dependence in the exponential in \eqref{eq:evaluationofaverage}. This quadratic dependence is related to assuming a constant-in-time large-scale random advection velocity. The model can be generalized to include temporal decorrelation of the large-scale sweeping velocity, which affects the shape of the frequency distribution.

In figure \ref{fig:wavenumberfrequencyspectra}(b) we test the prediction of the model for a fixed height $z/H \approx 0.154$. To this end, we obtain the wavenumber spectrum of the streamwise velocity component resolved with respect to the streamwise and spanwise wavenumbers from LES. Moreover, the values of the mean velocity and the variance of the streamwise and spanwise random advection (approximated by the variance of the streamwise and spanwise velocity components) have been obtained from LES. The joint $\bs k$-$\omega$ spectrum is then evaluated according to \eqref{eq:wavenumberfrequencyspectrum} before it is projected to the streamwise wavenumber and frequency. We note that this comparison does not involve any adjustable parameters. Figure \ref{fig:wavenumberfrequencyspectra} shows that the resulting spectrum captures the Doppler shift and broadening, including the non-vanishing Doppler broadening at low wavenumbers, quite well. On projecting the spectrum to the $k_1$-$\omega$-plane, as shown in figure \ref{fig:wavenumberfrequencyspectra}, an aliasing-like effect takes place: the spectral energy for a fixed $k_1$ and $\omega$ contains contributions from {\it all} $k_2$. As a consequence, the Doppler broadening associated with high $k_2$, for example, also contributes to low $k_1$. One of the immediate consequences is that the $k_1$-$\omega$ spectrum exhibits a non-vanishing Doppler broadening at $k_1=0$. This observation demonstrates that at least two spatial directions have to be taken into account in order to reproduce the ``aliasing'' by random sweeping effects (an early model focussing on the streamwise direction only was lacking this effect \citep{wilczek14jpc}).

The prediction of the model can be tested more quantitatively by considering normalized cuts through the $k_1$-$\omega$ spectrum, as presented in figure \ref{fig:wavenumberfrequencyspectra}(c). The  model captures the frequency distributions from LES quite well, and only slight deviations, especially in the amplitude of the Doppler broadening, are visible. These could be related to the various assumptions made in the derivation of the model, including the neglect of sweeping in the vertical direction.  Moreover, a slight asymmetry of the LES distributions can be noted, which is not captured by the model. The noticeable transition from Gaussian to non-Gaussian frequency distributions at decreasing wavenumbers, however, is captured quite accurately by the  model. It should be noted that from these data we did not find it necessary to introduce corrections to the local convection velocity. Further refinements could make use of the existing body of work that examines such possible corrections (see, e.g., the works by \cite{krogstad98pof} and \cite{alamo09jfm}).

\section{Full parameterization for $E_{11}(k_1,k_2,\omega;z)$}\label{sec:fullmodel}

The last section showed that the linear random advection model captures the frequency part of the $k_1$-$\omega$ spectrum. For application purposes a model for the wavenumber spectrum and parameterizations of the mean velocity and the random advection effects are desirable. This is the topic of this section.

The spectrum of the streamwise velocity component resolved with respect to the streamwise and spanwise wavenumbers already contains a considerable amount of information. Practical model parameterizations for joint streamwise-spanwise spectra are not firmly established, although a number
of experimental measurements have been presented, for example, by \cite{tomkins05jfm}. On the other hand, observations (see, e.g., \cite{marusic13jfm}) suggest that the streamwise wavenumber spectrum $E_{11}(k_1;z)$ for a range of heights in the logarithmic layer takes the form
 \begin{equation}\label{eq:streamwisemodelspectrum}
  E_{11}(k_1;z) =
  \begin{cases}
  \frac{C_1}{\kappa^{2/3}} u_*^2 H & k_1 \leq 1/H \\
  \frac{C_1}{\kappa^{2/3}} u_*^2 k_1^{-1} & 1/H < k_1 \leq 1/z \\
  C_1 \left(\frac{u_*^3}{\kappa z}\right)^{2/3} k_1^{-5/3} & k_1>1/z \, .
  \end{cases}
\end{equation}
Here, $C_1 = \frac{18}{55}C_{\mathrm{K}}$ is the Kolmogorov constant for the streamwise wavenumber spectrum related to the Kolmogorov constant $C_{\mathrm{K}} \approx 1.6$ from the energy spectrum function and $\kappa \approx 0.4$ is the von K\'arm\'an constant. The mean energy dissipation $\varepsilon$ is estimated as $\varepsilon = u_*^3/(\kappa z)$.  It should be noted that the functional form \eqref{eq:streamwisemodelspectrum} implies a logarithmic behavior for the variance as a function of the distance from the wall \citep{perry86jfm,davidson06pof}. Less is known about $E_{11}(k_2;z)$. We assume a $-5/3$ range for $k_2 > 1/z$ and a constant behavior for $k_2 \leq 1/z$, which appears to be a reasonable working hypothesis based on our limited-resolution LES data.

These considerations serve as guiding conditions for the parameterization of the joint wavenumber spectrum $E_{11}(\bs k;z)$. In the following $E_{11}(\bs k;z)$ is defined on the half-plane with positive $k_1$ as well as positive and negative $k_2$.

To model the high-wavenumber part of the model we assume isotropic turbulence obeying Kolmogorov scaling. The wavenumber spectrum in this range, in the following denoted as $E^{>}_{11}$, can then be related to the spectral energy tensor for homogeneous isotropic turbulence,
\begin{equation}
  \Phi_{ij}\big( \tilde{\bs k} \big) = \frac{E\big(\tilde k \big)}{4 \pi {\tilde k}^2} \left( \delta_{ij} - \frac{\tilde k_i \tilde k_j}{{\tilde k}^2} \right) \, ,
\end{equation}
where $\tilde{\bs k} = (k_1,k_2,k_3)^{\mathrm{T}}$ is the three-dimensional wavevector and $E\big(\tilde k \big)$ is the energy spectrum function for which we assume a Kolmogorov spectrum  $  E\big(\tilde k \big) = C_{\mathrm{K}} \varepsilon^{2/3} \tilde k^{-5/3}$. To obtain a simple analytical result we assume an infinitely extended inertial range, for which the wavenumber spectrum $E^{>}_{11}$ is obtained by
\begin{equation}\label{eq:smallscalespectrum}
  E^{>}_{11}(\bs k;z) = 2 \int \! \mathrm{d}k_3 \, \Phi_{11}\big( \tilde{\bs k} \big) = \int \! \mathrm{d}k_3 \, \frac{E\big(\tilde k \big)}{2 \pi {\tilde k}^2} \left( 1 - \frac{k_1^2 }{{\tilde k}^2} \right)   = \frac{\Gamma\left( \frac{1}{3} \right) C_{\mathrm{K}}}{5\sqrt{\pi}\Gamma\left( \frac{5}{6} \right)}  \varepsilon^{2/3} \left[ 1 - \frac{8}{11} \frac{k_1^2}{k^2} \right] k^{-8/3} \, ,
\end{equation}
with $\Gamma\left( \frac{1}{3} \right)/\left( 5\sqrt{\pi}\Gamma\left( \frac{5}{6} \right)\right)\approx 0.268$ (here, $\Gamma$ denotes the gamma function).
As can be expected, even for isotropic turbulence $E^{>}_{11}$ is not an isotropic function in the plane, i.e. it depends not only on $k$ but also on $k_1$.

At the large scales, wall-bounded flows are clearly anisotropic. To approximate the low-wavenumber transition between a $k_1^{-1}$ and a constant spectrum, we use a power-law blending according to
\begin{equation}\label{eq:largescalespectrum}
  E^{<}_{11}(\bs k;z) = D(z) z u_*^2 \left[ \left({1}/{H}\right)^{\beta}  + k_1^{\beta} \right]^{-1/\beta} \, ,
\end{equation}
with a non-dimensional height-dependent amplitude $D(z)$, which is determined numerically such that the fluctuation variance of the model spectrum matches the one obtained from the log law \eqref{eq:loglawfluctuations} below. The exponent $\beta=4$ is chosen by empirical fitting.

\begin{figure}
\begin{center}
  \topinset{(a) \hspace{3.3cm} (b)}{\includegraphics[width=0.7\textwidth]{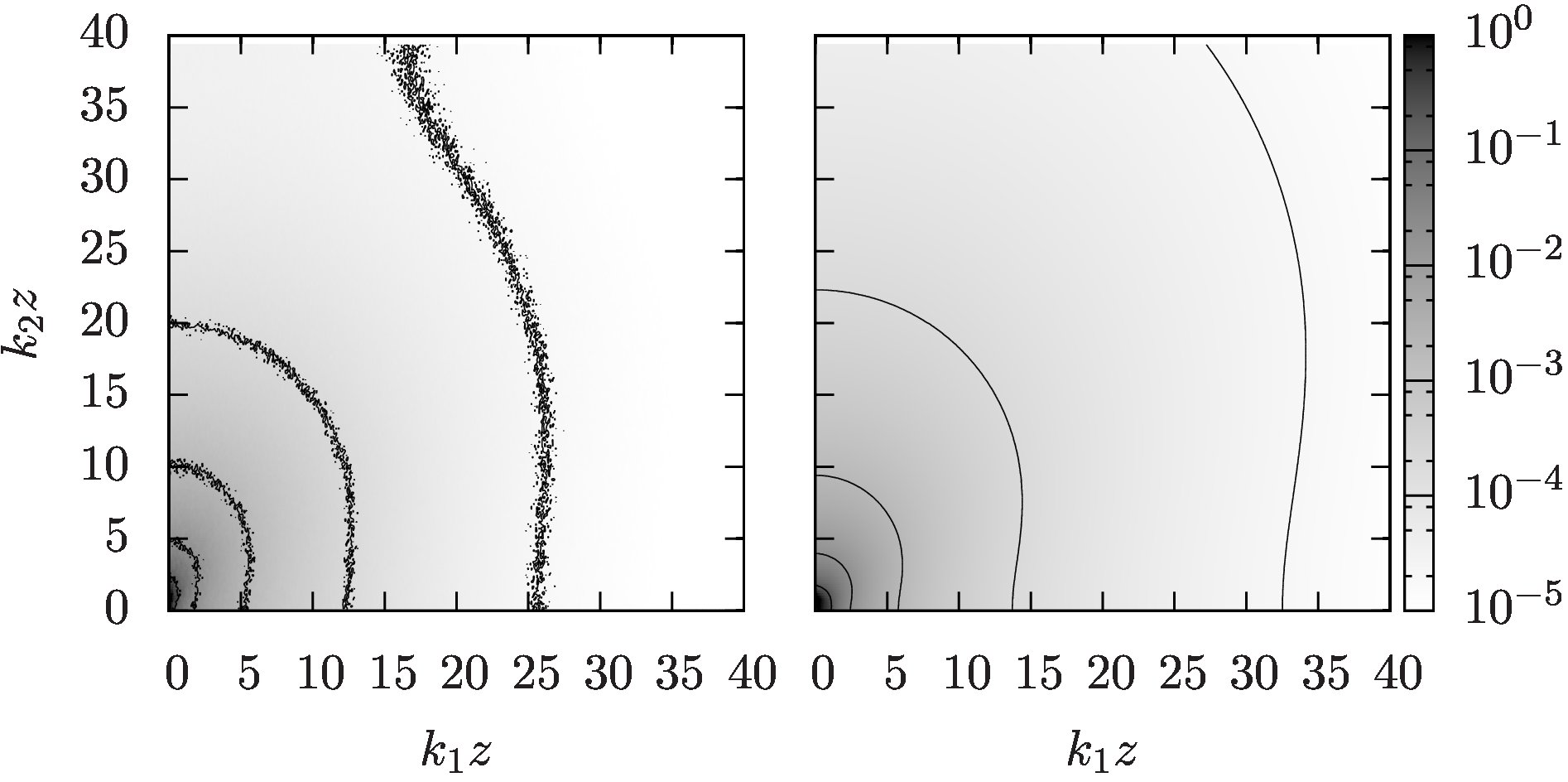}}{0.4cm}{-1.4cm}
  \topinset{(c) \hspace{3.3cm} (d)}{\includegraphics[width=0.7\textwidth]{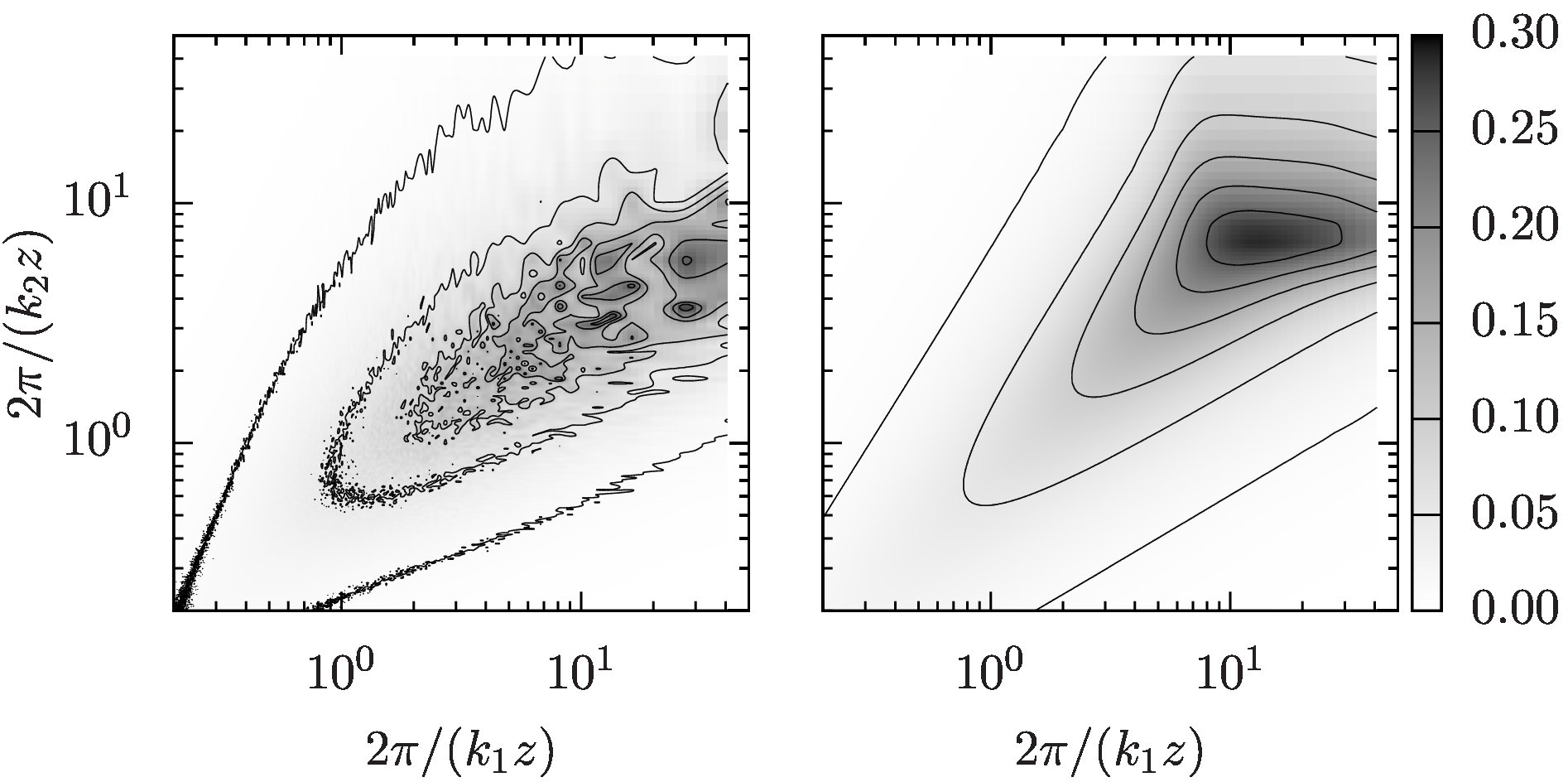}}{0.4cm}{-1.4cm}
\end{center}
\caption{(a) Spectrum of the streamwise velocity component from LES at $z/H \approx 0.154 $, resolved with respect to the streamwise and spanwise wavenumbers. (b) Model spectrum \eqref{eq:streamspanspecfullmodel}. (c,d) The same as (a,b) but in premultiplied representation.  }
\label{fig:streamspanspectrum}
\end{figure}

Between these two regimes we smoothly blend with a sigmoidal function $\theta_{\alpha}(x) = \left( \tanh[ \alpha \log(x) ] +1 \right)/2$, where $\alpha$ controls the steepness of the step. We choose $\alpha=4$ in the following. Combining these individual pieces, our model wavenumber spectrum takes the form
\begin{equation}\label{eq:streamspanspecfullmodel}
  E_{11}(\bs k ; z) = \left[1 - \theta_{\alpha}\left( kz \right)\right] E^{<}_{11}(\bs k ; z) + \theta_{\alpha}\left( kz \right) E^{>}_{11}(\bs k ; z)  \, .
\end{equation}
Figure \ref{fig:streamspanspectrum} shows a comparison of the streamwise-spanwise spectrum from LES data and the model spectrum at $z/H\approx 0.154$. While the qualitative features compare quite well, the LES spectrum, as compared with the model, exhibits a more pronounced large-scale anisotropy and a faster decay at large wavenumbers. This means that the current model wavenumber spectrum leaves room for improvement with respect to capturing the statistical features of the large scales as well as of the turbulent fluctuations on smaller scales. We furthermore note that the model wavenumber spectrum \eqref{eq:streamspanspecfullmodel} only approximately reduces to the one-dimensional spectrum \eqref{eq:streamwisemodelspectrum}. In particular, the low-wavenumber range deviates slightly from a clean $-1$ range (not shown). This is related to the circular blending of the two contributions and aliasing of the $-5/3$ range of the model spectrum to the low-$k_1$ range.

\begin{figure}
\begin{center}
  \topinset{(a) \hspace{8.1cm} (b)}{\includegraphics[width=1.0\textwidth]{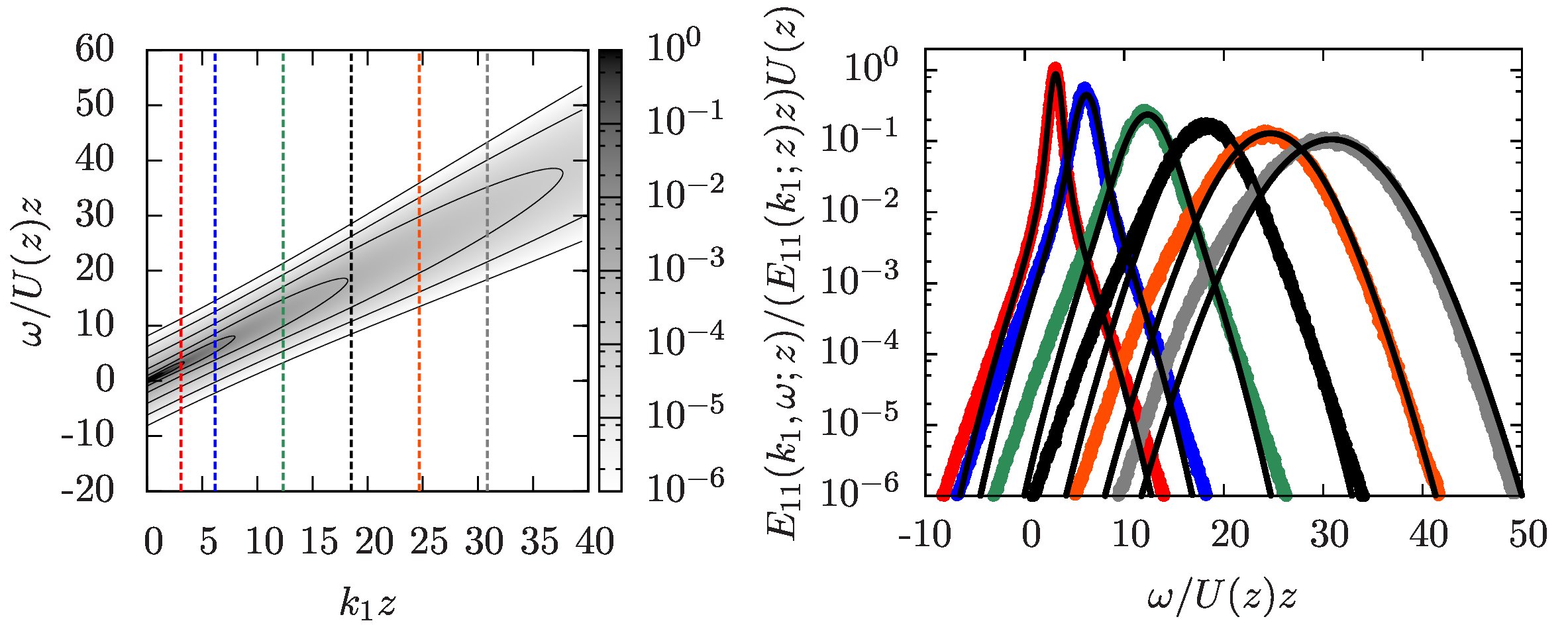}}{0.6cm}{1.5cm}
\end{center}
\caption{(a) Fully modeled $k_1$-$\omega$ spectrum of the streamwise velocity component at $z/H \approx 0.154$. From the log laws \eqref{eq:loglawmean} and \eqref{eq:loglawfluctuations}, we obtain $U \approx 18.4 \, u_*$ and $\langle v_1^2 \rangle \approx 4.21 \, u_*^2$. (b) Frequency distributions for the same model spectrum at various wavenumbers (black lines) compared with the normalized cuts through the wavenumber frequency spectrum from LES (colors).}
\label{fig:spectrafullmodel}
\end{figure}

As seen in section \ref{sec:linadvectioneq}, the temporal part of the model includes a Gaussian frequency distribution with a mean $\mu(z) =  k_1 U(z)$ and a variance $\sigma^2(z) = \langle (\bs v \cdot \bs k)^2 \rangle$. We restrict ourselves to the logarithmic layer, so that for a rough boundary layer with a roughness length $z_0$ the mean velocity profile is well approximated by 
\begin{equation}\label{eq:loglawmean}
  U(z) = \frac{u_*}{\kappa} \log\left( \frac{z}{z_0} \right) \, .
\end{equation}
The variance of the frequency distribution depends on both components of the random advection velocity,
\begin{equation}\label{eq:loglawvariance}
  \sigma^2(z) = \left\langle (\bs v \cdot \bs k)^2 \right\rangle = \left\langle v_1^2 \right\rangle k_1^2 + \left\langle v_2^2 \right\rangle k_2^2 \, .
\end{equation}
The cross-term vanishes because $\left\langle v_1v_2 \right\rangle=0$. Recently, a logarithmic dependence of the streamwise velocity fluctuation has also been established \citep{marusic03pof,hultmark12prl,marusic13jfm}, and since the major contribution to the total variance comes from the large scales, we model the variance as
\begin{equation}\label{eq:loglawfluctuations}
  \left\langle v_1^2 \right\rangle = u_*^2 \left[ B - A\log\left( \frac{z}{H} \right) \right] .
\end{equation}
Here, $A$ is the ``Perry-Townsend" constant and $B$ is a non-universal constant depending on the specific flow. For our evaluations we obtain $A = C_1/\kappa^{2/3} \approx 0.965$ and $B = 5C_1/(2\kappa^{2/3}) \approx 2.41$, which can be derived from the model spectrum \eqref{eq:streamwisemodelspectrum} for the streamwise velocity component. For an extensive discussion on the logarithmic behavior of streamwise velocity fluctuations and its relation to a spectral budget model we refer to \cite{banerjee13pof}. We note, however, that larger values for $A$ (close to $A \approx 1.25$) have also been reported in the literature (see, e.g., \cite{marusic13jfm}). The values of $B$ vary across different numerical data sets and experiments; however, they are not expected to be universal. To simplify the model parameterization, we assume that $\langle v_2^2 \rangle$ is proportional to $\langle v_1^2 \rangle$ (which is an approximation), such that the variance of the frequency distribution takes the form
\begin{equation} \label{eq:totalvariance}
  \sigma^2(z) = \left\langle v_1^2 \right\rangle \left [ k_1^2 + C k_2^2 \right] \, .
\end{equation}
Our LES simulations suggest a typical value of $C \approx 0.41$ in the range of heights under consideration.

The entire model for $E_{11}(\bs k,\omega;z)$  is thus fully specified by evaluating the prediction based on the linear advection equation \eqref{eq:wavenumberfrequencyspectrum} in conjunction with \eqref{eq:smallscalespectrum} -- \eqref{eq:totalvariance} and the quoted numerical parameters.
Figure \ref{fig:spectrafullmodel}(a) shows the streamwise $k_1$-$\omega$ spectrum for $z/H \approx 0.154$ from the full model. The integration over the spanwise wavenumber, necessary to obtain the projected streamwise $k_1$-$\omega$ spectrum, was carried out numerically. To this end the analytical model was discretized on a grid matching the one of the reference LES data presented here. The spectrum agrees well with the LES data. This is also confirmed by a plot of the normalized cuts, see panel (b) of the same figure. The good agreement also rests on the fact that the model spectrum is restricted to the same wavenumber range as the LES data. If we extend the model spectrum to higher wavenumbers, aliasing effects contribute to additional Doppler broadening into lower frequencies. This also implies that LES, compared with an infeasible DNS, underestimates the frequency broadening in the projected streamwise $k_1$-$\omega$ spectrum.

\section{Summary}

We have evaluated the $\bs k$-$\omega$ spectrum of the streamwise velocity component from LES data in the logarithmic region of  a wall-bounded flow. The frequency distribution exhibits a Doppler shift induced by mean flow advection as well as a Doppler broadening, which to leading order is caused by large-scale random advection effects. These effects were then used as the main ingredients for a model for the $\bs k$-$\omega$ spectrum, leading to the prediction that the joint $\bs k$-$\omega$ spectrum can be written as a product of the wavenumber spectrum with a Gaussian frequency distribution. We find that the model predictions agree well with the LES data.

In order to obtain an analytically tractable model for the entire $\bs k$-$\omega$ spectrum, we  proposed a model parameterization in which the Doppler shift and broadening in the frequency distribution were parameterized by means of log laws for the mean velocity and the variance. Along with a model for the wavenumber part of the spectrum, this full model parameterization also exhibits good agreement with the LES data.

Based on the simple idea of large-scale advection of the small-scale fluctuations, the model can be generalized to include additional effects like shear (see, for example, \citet{mann94jfm} for such an approach). It can also be combined with alternative parameterizations of the spectral energy tensor. Furthermore, the analytical model for the $\bs k$-$\omega$ spectrum can be extended to applications such as characterizing velocity fluctuations in wind farms interacting with the turbulent atmospheric boundary layer, which is the topic of ongoing work.
\\
\\
{\it Acknowledgments:} 
M.W. was supported by DFG funding WI 3544/2-1 and WI 3544/3-1, R.J.A.M.S. by the `Fellowships for Young Energy Scientists' (YES!) of FOM and C.M. by US National Science Foundation grant \#IIA-1243482 (the WINDINSPIRE project). Computations were performed with SURFsara resources, i.e. the Cartesius and Lisa clusters. This work was also supported by the use of the Extreme Science and Engineering Discovery Environment (XSEDE), which is supported by National Science Foundation grant number OCI-1053575.
\\
\\


\end{document}